\documentclass{iopart}

\def\erf{{\rm erf}}
\def\sgn{{\rm sgn}}
\def\lda{\left\langle\!\left\langle}
\def\rda{\right\rangle\!\right\rangle}
\def\lb{\left}
\def\rb{\right}
\newcommand{\fr}[2]{\frac{#1}{#2}}
\def\bea{\begin{eqnarray}}
\def\eea{\end{eqnarray}}
\def\nn{\nonumber}
\def\be{\begin{equation}}
\def\ee{\end{equation}}

\usepackage{amssymb}
\usepackage[english]{babel}
\usepackage{color,epsfig}
\usepackage{graphicx}

\begin{document}

\title[Majority game]{Statistical mechanics of the majority game}
\author{P Koz{\l}owski$\dagger\ddagger$ and M Marsili$\dagger$}
\address{$\dagger$ Abdus Salam International Centre for Theoretical Physics,\\
 Strada Costiera 11, 34014 Trieste}
\address{$\ddagger$ Istituto Nazionale per la Fisica della Materia (INFM),\\ 
Unit\'a Trieste-SISSA, Via Beirut 2-4, 34014 Trieste}
\eads{\mailto{kozl@ictp.trieste.it}, \mailto{marsili@ictp.trieste.it}}

\begin{abstract}
The majority game, modelling a system of heterogeneous agents trying
to behave in a similar way, is introduced and studied using methods of
statistical mechanics. The stationary states of the game are given by
the (local) minima of a particular Hopfield like hamiltonian. On the
basis of a replica symmetric calculations, we draw the phase diagram,
which contains the analog of a retrieval phase. The number
of metastable states is estimated using the annealed
approximation. The results are confronted with extensive numerical
simulations.
\end{abstract}

\section{Introduction}

The cooperative processes taking place in complex systems of agents
interacting with each other, which were up to quite recently studied
mainly by human sciences, became also an interesting object of
research for physicists.  Despite the tremendous
complexity of such systems and the presence of the unpredictable
ingredients related to human free will, some of the statistical
regularities which characterise their collective behaviour can be studied
using models and techniques developed in the field of statistical
physics.

As usually in such situations one builds very abstract theoretical
models which have a much wider applicability. The so called minority
game, for example, was proposed as a formalisation of the El-Farol bar
problem \cite{elfarol}, in order to capture the essential features of
traders' interaction in a stock exchange market \cite{minority}. But
the same model can describe a more general situation where many agents
compete for the exploitation of a number of scarce resources
\cite{MMRT}.

This seemingly simple model turns out to exhibit a surprisingly rich
variety of complex behaviours which were studied thoroughly using
various methods \cite{matteo_stat,coolen,matteo_con}. The statistical
mechanics approach to this sort of  models is particularly interesting as it
reveals aspects of cooperative phenomena which may be qualitatively
different from those studied in physics.

The minority game is based on the assumption that agents prefer to
avoid crowds. Hence they tend to use those strategies which let them
be as often as possible in the minority. This is only one of many
possible types of interactions. It seems natural to ask what happens
under the reversed (majority) rule, favouring those strategies
which allow the agents to stay on the side of the majority.

In the economic interpretation of this sort of models, it has been
shown \cite{M01} that the minority rule describes so-called
``fundamental'' or ``contrarian'' traders in a financial market. These
are agents who believe that market prices are close to an equilibrium
and hence expect that price fluctuations tend to generate price
changes towards the equilibrium value. The majority rule is instead
appropriate for trend followers, whose behaviour is thought to be
responsible for the so called ``bubbles'' -- buy rushes leading to
price increases well beyond those which would be justified by an
economic evaluation. The majority mechanism is self-reinforcing as it
generates self-fulfilling prophecies: if the agents expect that the
price will rise (fall), the majority of them will buy (sell) which
will actually make the price rise (fall).

Before focusing on the majority game, let us mention that the
competition between majority and minority players has been studied in
Ref. \cite{M01} in the simplest setting and in Ref. \cite{demartino}
in its full complexity.

From a wider perspective, the majority game describes a situation
where the profit of agents increases with the number of agents acting
in the same way. Conformity effects of this type are evident in the
spreading of fashions. A further example may be that of a shop which
lowers the prices as the number of customers increases, or of a product
which becomes cheaper the more it is popular. This mechanism, which
goes under the name of {\em increasing returns} in economics, lies at
the heart of quite interesting aggregation phenomena -- for example
the emergence of cities and economic districts such as Silicon Valley
and Hollywood \cite{Krugman}.

As we shall see the study of majority game leads to the analysis of
models which are very similar to attractor neural networks, in
particular to the Hopfield model \cite{Hopfield}. In brief, agents'
learning dynamics is different from Glauber dynamics, but the energy
landscape where it takes place is the same. It turns out that
aggregation in the majority game is the same phenomenon as memory
retrieval in the Hopfield model. Hence the physics of neural networks
tells us a lot about the behaviour of the majority game. On the other
hand, this study also provides new results on the physics of neural
networks by probing the energy landscapes such as that of the Hopfield
model with a different type of dynamics.

Our work is based on a statistical mechanics approach to the
stationary states of the majority game and its results are verified by numerical
simulations.
The paper is organised as follows: In
the next section we introduce the model and in section 3 we discuss
its stationary states -- we show that they can be
identified with the local minima of the Hopfield type hamiltonian. Section
4 deals with the calculation of free energy and the construction of
the phase diagram. Given that stationary states are selected in a
dynamical way and not according to a Boltzmann weight, we compare our
results with extensive numerical simulations. In order to clarify the
dynamical behaviour of the model, we discuss the number of stationary
states in section 5. We conclude with a summary of the main results
and a discussion of their implications.

\section{The definition of the model}

We consider a system consisting of $N$ agents interacting at discrete
time intervals (at each round of the game). The interaction takes
place through the action, concerning $p$ objects or resources, which
each agent undertakes. The actions are determined by one of $r$
strategies which are chosen randomly and independently for each of the
agents at the beginning of the game. In the course of the game the
agents can change their actions only by changing their strategies,
i.e. choosing one of the $r$ predefined ones (for each agent). We
denote by $a_{is}^\mu$ the action taken by agent $i=1,\ldots,N$
concerning resource $\mu=1,\ldots,p$ when he/she adopts strategy
$s=1,\ldots,r$. We consider here the case of binary actions
$a_{is}^\mu=\pm 1$. The specific values of $a_{is}^\mu$ -- the
realisation of quenched disorder -- are drawn at random from some
distribution.
Thus a strategy is a binary
vector, which can be interpreted for instance as a list of actions to
be undertaken concerning each of the objects.

Agent $i$ chooses the strategy used in the next round of the game on
the basis of the performances of his/her strategies in the previous
runs, which are measured by a score functions $u_{is}(t)$. The agents
choose the strategy 
\begin{equation}
s_i(t)={\rm arg}\max_s u_{is}(t)~\label{prob}
\end{equation}
with the highest score and undertake the profile of actions $a_{is_i(t)}^\mu$
$\mu=1,...,p$ ($i=1,...,N$).
The agents know about the actions of others only through
the cumulative actions:

\begin{equation}
A^\mu(t)=\sum_{i=1}^N a_{is_i(t)}^\mu~,
~~~~~~\mu=1,\ldots,p 
\end{equation}
which are used to update the score functions:

\begin{equation}
u_{is}(t+1)=u_{is}(t)+\frac{\epsilon}{p}\sum_{\mu=1}^p a_{is}^\mu \left[A^\mu(t)
-\eta(a_{is_i(t)}^\mu-a_{is}^\mu)\right]~,\label{majrule}
\end{equation}
where $\epsilon>0$ and $\eta\in [0,1]$ are constants.

Let us first discuss this dynamics for $\eta=0$. Note that $A^\mu(t)$
has the same sign as the action undertaken by the majority concerning
object $\mu$. Then eq. (\ref{majrule}) implies that those strategies
prescribing an action aligned with the majority are rewarded. In other
words, by this learning dynamics, agents strive to find that strategy
which puts them in the majority. Because of the averaging over $\mu$
this rule is called a batch version of the majority game. The on-line
version, where a value $\mu(t)$ is randomly drawn at each time and
agents update the scores depending on $A^{\mu(t)}(t)$, will be
discussed in the concluding section.

With parameter $0<\eta<1$ one can change the degree to which the
agents take into account the influence of their own actions on the
cumulative quantity $A^\mu(t)$ (see \cite{matteo_eta}). In particular
the case $\eta=1$ describes agents who are learning to respond
optimally to the behaviour of the others. Indeed for $\eta=1$
eq. (\ref{majrule}) computes the correct value of $A^\mu(t)$ if agent
$i$ had actually played strategy $s$. This is what game theory assumes
a rational player should do, so the stationary states of the game for
$\eta=1$ are Nash equilibria (i.e. those states where each agent
takes the optimal strategy, given the strategy of
others~\cite{GameTheory}). Like in the Minority game \cite{matteo_stat}, 
in spite of the fact that
$A^\mu(t)\sim\sqrt{N}$ is much larger than $a_{is}^\mu\sim O(1)$, the
$\eta$ term is not negligible.

The new, updated payoff functions are used to determine the action in
the next time step through eq. (\ref{prob}).

Note that $A^\mu(t)$ is the difference between the size of the two
groups of agents who undertake opposite actions $a^\mu_{is_i}=+1$ or
$-1$. If agents do not interact and the actions $+1$ and $-1$ are
equivalent, it is obvious that $A^\mu(t)\sim \sqrt{N}$.  We shall pay
special attention, in what follows, to the possibility that, when turning
on the interaction, a {\em macroscopic difference} $A^\mu(t) \propto
N$ may emerge, for some value of $\mu$.

\subsection{$r=2$ case}

In this paper we will focus on the case where $r=2$. We allow for a
correlation of the two strategies of the same agent by introducing a
parameter 
\begin{equation}
g=P(a_{i+}^\mu=a_{i-}^\mu)\label{defg}~,
\end{equation}
with $P(a_{i+}^\mu=1)=P(a_{i+}^\mu=-1)=\frac{1}{2}$. 

Instead of keeping track of the two payoff functions it is enough to
consider their difference:

\begin{equation}
y_i=\frac{u_{i+}-u_{i-}}{2}
\end{equation}

\section{Stationary states}

Taking the average over the stationary state distribution, which we
denote by $\langle ...\rangle$, of the dynamical equation of $y_i(t)$,
we get:

\begin{equation}
 v_i\equiv\fr{\langle y_i(t+1)-y_i(t)\rangle}{\epsilon}
 =\overline{\xi_i\Omega}+\sum_{j=1}^N\overline{\xi_i\xi_j}m_j
 -\eta\overline{\xi_i^2}m_i~, \label{dy}
\end{equation}

\noindent where we used a standard notation:

\[ 
\fl m_i=\langle {\rm sign}\, y_i
\rangle,~~~~\xi_i^\mu=\frac{a_{i+}^\mu-a_{i-}^\mu}{2},~~~~
\Omega^\mu=\sum_{j=1}^N\frac{a_{i+}^\mu+a_{i-}^\mu}{2},~~~~
\overline{x}=\frac{1}{p}\sum_{\mu=1}^px^\mu
\]

We expect that all the
agents in the long time limit are frozen, i.e. they do not change
their strategies. Indeed, exactly as in the case of the minority game,
it is easy to check that the stationary states correspond to the
minima of
\begin{eqnarray}
{\cal
H}_\eta&=&-\fr{1}{2}\overline{A^2}+\frac{\eta}{2}\sum_{i}
\overline{\xi_i^2}m_i^2\nn
\\
&=&-\fr{1}{2}\sum_{i,j}\overline{\xi_i\xi_j}m_im_j-\sum_i
\overline{\Omega\xi_i}m_i
-\fr{1}{2}\overline{\lb(\Omega^\mu\rb)^2}+
\frac{\eta}{2}\sum_{i}\overline{\xi_i^2}m_i^2
\label{ham}~.
\end{eqnarray}
The argument starts by observing that if $v_i\neq 0$, then $y_i\to
\pm\infty$, depending on the sign of $v_i$, and hence $m_i={\rm sign}\,
v_i$. Only if $v_i=0$ we can have $m_i\neq \pm 1$. Then one observes
that $v_i=-\frac{\partial {\cal H}_\eta}{\partial m_i}$ so these conditions
are equivalent to the conditions for the minima of ${\cal H}_\eta$. But it
is evident, from the form of ${\cal H}_\eta$, that its minima lie only at
the corners of the hypercube $[-1,1]^N$.

The conclusion that stationary states correspond to the minima of
${\cal H}_\eta$ is straightforward from the dynamical equations in the
limit $\epsilon\to 0$. Then one can introduce a rescaled continuous
time $\tau=\epsilon t$ and verify that ${\cal H}_\eta$ is a Lyapunov
function of the continuum time dynamics.

\begin{figure}[h]
\includegraphics[width=10cm,height=6cm,angle=0]{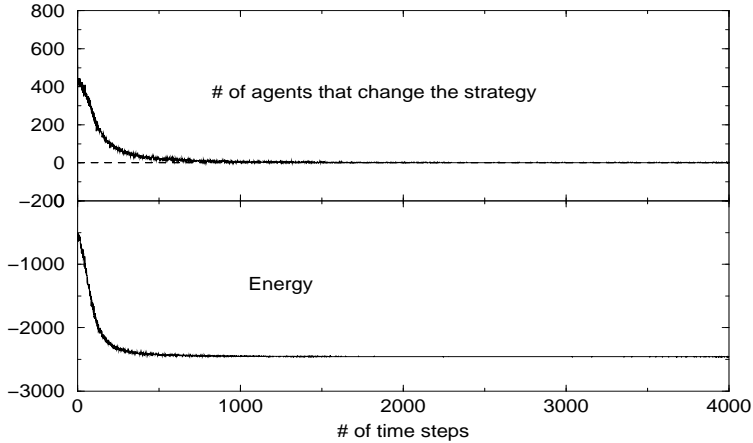}
\caption{Simulations of the model for $\alpha=0.2$, $N=1000$, $\eta=0$
and random initial conditions.}
\label{sim1}
\end{figure}

Fig. \ref{sim1} shows that numerical simulations fully confirm the
above picture. Note in particular that while the initial stages of the
dynamics are somewhat noisy, fluctuations are negligible in the long
time limit.

Strictly speaking there is no stationary state in terms of the
variables $y_i$ as they diverge to $\pm \infty$. The term {\em stationary
state} refers to the variables $m_i$ which take well defined values in
the limit $t\to\infty$.

We remark that the ${\cal H}_0$ is simply related to the predictability
$H=-2{\cal H}_0$ introduced in the minority
game~\cite{matteo_con}. Since $m_i=\pm 1$ for all $i$, $H$ is also
equal to the volatility $\sigma^2=\overline{A^2}$. Hence agents in the
majority game strive to maximise $H$ whereas agents in the minority
game minimise it.

${\cal H}_\eta$ is very similar to the Hamiltonian of the Hopfield model
known from the theory of neural networks \cite{nn}. The only
differences are: the scaling with $1/p$ instead of $1/N$, the presence
of the random field and the fact that the patterns $\xi_i^\mu$ can
take three values $0,\pm 1$ instead of only two $\pm 1$. 
By analogy with physics we shall call ${\cal H}_\eta$ energy. We notice
that for $g=0$ one obtains the ``pure'' Hopfield model (with different
rescaling), whereas for $g=1/2$ one has a majority game with
independently chosen strategies.

The stationary state values of $m_i$ satisfy the equations:
\begin{eqnarray}
m_i=\sgn\left(\overline{\xi_i\Omega}+\sum_{j=1}^N\overline{\xi_i\xi_j}m_j
-\eta\overline{\xi_i^2} m_i\right)\label{stab}
\end{eqnarray}
It is clear that any configuration ${\cal C}=\{m_i\}$ which is a
solution of these equations for some value of $\eta\in [0,1]$ will
also be a solution for all $\eta'<\eta$. Hence the set ${\cal S}_\eta$
of stationary states satisfies the property ${\cal S}_\eta\subset
{\cal S}_{\eta'}$ for $\eta'<\eta$ and, in particular, ${\cal
S}_1\subset {\cal S}_\eta$ for all $\eta<1$. It is also easy to see
that the state with minimal value of ${\cal H}_\eta$ lies in ${\cal
S}_1$ for all $\eta\in [0,1]$. This shows that Nash equilibria are
stationary state of the majority game for all values of $\eta$, but
the converse is not true.
 
In the remaining part of this article we will try to determine the
nature of the stationary states of the majority game and calculate
the number of such states.  In the next section we will employ the replica
method to analyse the thermodynamics (for $T=0$) of the model defined
by the Hamiltonian (\ref{ham}). Then, using the stability relation
(\ref{stab}) we will calculate the number of stationary states in the
annealed approximation as a function of various parameters .

\section{Replica approach to thermodynamics}

In the light of the previous considerations, it is clear that
stationary states of the majority game for all values of $\eta\in
[0,1]$ are determined by the minima of ${\cal H}_\eta$ lying at the
corners of the hypercube $[-1,1]^N$. Note that when we restrict
ourselves to $m_i=\pm 1$ the $\eta$ term in ${\cal H}_\eta$ becomes an
irrelevant constant. This is why all the results in this section do not
depend on $\eta$.

As usual, we build a partition function corresponding to the
Hamiltonian (\ref{ham}), introducing an inverse temperature
$\beta$. Using the replica method \cite{replica}, we perform the
average over quenched disorder and obtain the free energy density
function. Within the replica symmetric ansatz, the latter depends on
the Edward Anderson order parameter $q$, the overlap $b$ with pattern
$\xi^1$ and the residual overlap $r$. In the limit $\beta\to\infty$ we
find $q\to 1$ with
\[
\chi=\lim_{\beta\to\infty} \beta(1-q)
\]
finite and the free energy takes the form: 

\bea
f&=&\fr{1}{2}\fr{b^2}{\alpha}+\alpha\chi r
-\fr{\alpha}{2}\fr{1}{\alpha-(1-g)\chi}
-2\sqrt{\fr{\alpha r}{\pi}}\left[g+(1-g)e^{-b^2/(4\alpha^3 r)}\right]\nn \\ 
&~& -\frac{(1-g)b}{\alpha}{\rm
  erf}\left(\frac{b}{2\alpha^{3/2}\sqrt{r}}\right).
\eea

The parameters $b$, $r$ and $\chi$ should take the values optimising
the free energy. We remark that the overlap $b$ corresponds to the
equilibrium value of $A^1/N$.  The capacity $\alpha=p/N$, the reversed
temperature $\beta$ and $g$ defined by (\ref{defg}) are the
parameters. 

\begin{figure}[h]
\includegraphics[width=12cm,height=7cm,angle=0]{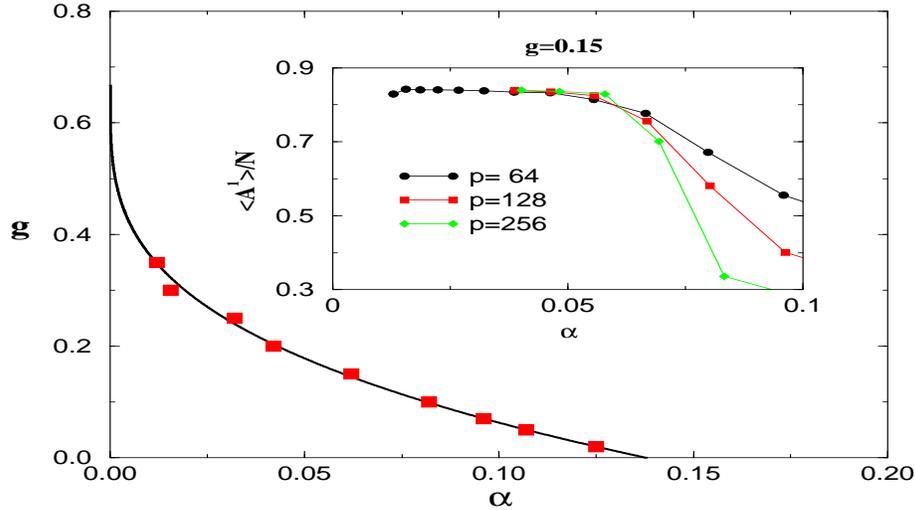}
\caption{The phase diagram computed by the replica symmetric
  approach described in the text (full line). The points are numerical
  estimate of the phase boundary $\alpha_c(g)$ obtained from the
  crossing of the maximal overlap for different system sizes, as shown
  in the inset ($g=0.15$, $\eta=1$, $p=64,~128$ and $256$, maximal overlap
  initial conditions).}\label{phdiag}
\end{figure}

The analysis of the solutions to the saddle point equations 

\be
\frac{\partial f}{\partial b}=0~,~~~\frac{\partial f}{\partial r}=0~,
~~~\frac{\partial f}{\partial \chi}=0
\label{fixed}
\ee

\noindent 
allows us to draw the phase diagram (see fig. \ref{phdiag}). 

One can distinguish two phases: a spin glass phase ($b=0$) and a
retrieval phase, where besides a spin glass solution there exists also a
retrieval solution ($b\neq 0$).

\begin{figure}[h]
\includegraphics[width=11cm,height=7cm,angle=0]{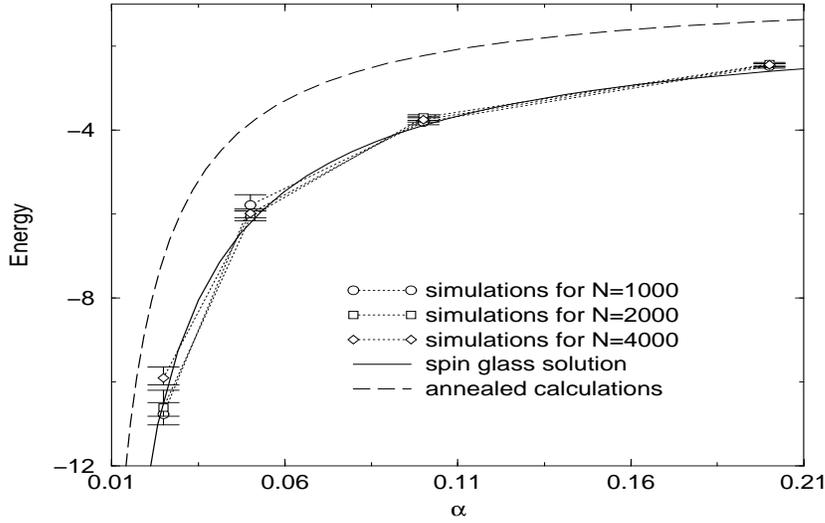}
\caption{Comparison of energy profiles obtained for $g=0.5$ and
$\eta=0$.}\label{encom}
\end{figure}

A simple calculation shows that for the spin glass solution

\[
\fr{\overline{A^2}}{N}=\frac{-2{\cal H}_0}{N} \cong
\left(1+\sqrt{\fr{2(1-g)}{\pi \alpha}}\right)^2 
\]
within the replica symmetric ansatz (see Fig. \ref{encom}). The
solution with $b\neq 0$, which using the neural network nomenclature,
we call retrieval, is conveniently described in terms of the
parameter $x=b/(2\alpha\sqrt{\alpha r})$ \cite{ags87}, which satisfies
the equation
\begin{equation}
x=\frac{(1-g){\rm
    erf}(x)}{\sqrt{2\alpha(1-g)}+2\pi^{-1/2}(1-g)[g+(1-g)e^{-x^2}]}. 
\label{x}
\end{equation}
For $g\le 2/3$, two non-zero solutions for $x(\alpha)$ exist up to a
critical value $\alpha_c(g)$, but only one of them represents a
thermodynamically stable state. Of course for $g\rightarrow 0$ we find
that the spin glass and retrieval solutions of (\ref{fixed}) converge
to the solutions found for the Hopfield model \cite{ags87}. The phase
separation line $\alpha_c(g)$ in fig. (\ref{phdiag}) smoothly
approaches the $\alpha=0$ axis $\alpha_c\simeq
\frac{75}{2\pi}\left(\frac{2}{3}-g\right)^4$ when $g\to 2/3$.

In the retrieval phase the dynamics can, depending on the initial
conditions, end up in one of the two qualitatively different sorts of
attractors. In the majority game language the retrieval corresponds
to the macroscopic value of $A^\mu\sim O(N)$ for some $\mu$ (say $1$),
whereas $A^\mu\sim\sqrt{N}$ for the remaining values of $\mu=2,...,p$.
In the spin glass phase, $A^\mu\sim\sqrt{N}$ for all $\mu$.

To confirm these analytical results we have performed extensive
numerical simulations using the dynamical definition of the model
(\ref{prob}, \ref{majrule}). This is important because the above
calculation is based on the Boltzmann weight and it focuses on the
lowest energy minima. There is, however, no guarantee that the dynamics
of the majority game selects the minima with the lowest energy.

Direct iteration of the dynamics is slow because in the late stages
the time interval between individual spin flips $s_i\to -s_i$
becomes very large. A much more efficient algorithm is possible in the
continuum dynamics for $\epsilon\to 0$ because then one can integrate
easily the dynamics between two consecutive spin flips.
Clearly when $t\gg
1$ the continuum time dynamics coincides with the discrete time batch
dynamics ($\epsilon=1$). Indeed no noticeable difference between
simulations with $\epsilon=1$ and $\epsilon \to 0$ was found in the
typical properties of stationary states.

The comparison of the energies obtained using different methods for
few values of $\alpha$ in the spin glass phase is presented in
fig. \ref{encom}.

The estimate of the phase separation line coming from
the simulations is plotted in fig. \ref{phdiag}. Each point was
obtained from the crossing of the curves $A^1/N$ vs $\alpha$ for
different system sizes, as shown in the inset. There is a good
agreement with the analytical results.  The static results are $\eta$
independent, but simulations clearly show that this parameter
plays an important role in the dynamics.

\begin{figure}[h]
\includegraphics[width=12cm,height=7cm,angle=0]{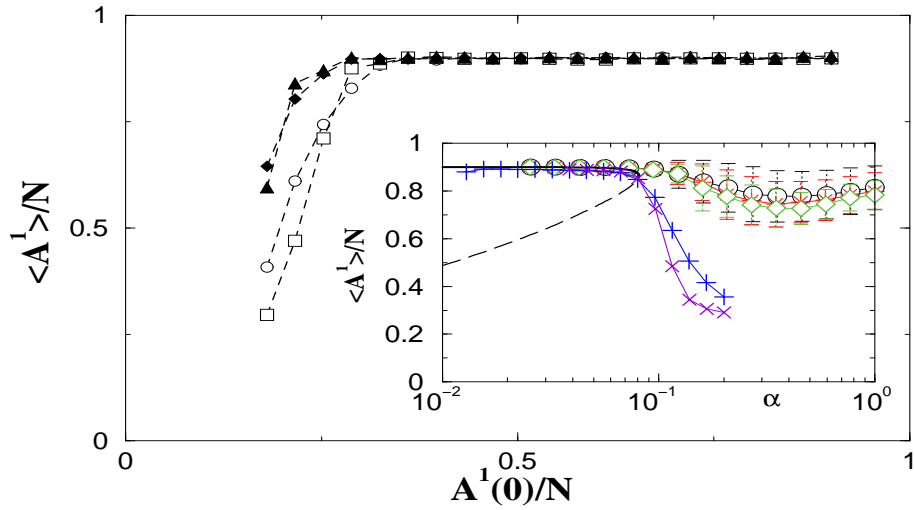}
\caption{Overlap $A^1/N$ in the stationary state as a function of
initial overlap at $t=0$ in the retrieval phase ($\alpha=0.05$ and
$g=0.1$) for $\eta=0$ ($\blacklozenge$ and $\blacktriangle$) 
and $1$ (\opencircle and \opensquare). Two system sizes ($p=64$ and $128$) are
shown in order to appreciate finite size effects. Inset: Overlap as a
function of $\alpha$ for $\eta=0$ (\opencircle, $*$ and \opendiamond) and
$\eta=1$ ($+$ and $\times$) for maximal initial overlap.  The stable
(solid line) and unstable (dashed) solutions of saddle point equations
are also shown.}\label{corrup}
\end{figure}

Fig. \ref{corrup} shows that retrieval states are indeed attractors in
the retrieval phase. Even when starting from initial conditions
which have only a partial overlap with pattern $\mu=1$, the dynamics
converges to the retrieval state both for $\eta=0$ and $1$. Actually,
retrieval is enhanced when $\eta=0$.

The inset of fig. \ref{corrup} shows how the overlap $A^1/N$ depends
on $\alpha$, in simulations with maximal initial overlap
$A^1(t=0)/N$. While for large $\alpha$ the overlap essentially vanishes (in the
limit $N\to\infty$) 
if $\eta=1$, $A^1/N$ attains a relatively large value for $\eta=0$. 
The error bars for $\eta=0$ indicate that the distribution of the overlap 
is very broad for intermediate values of $\alpha$; the overlap in a 
particular run can converge to any value in the interval $[0,1-g]$. 

\begin{figure}[h]
\includegraphics[width=12cm,height=7cm,angle=0]{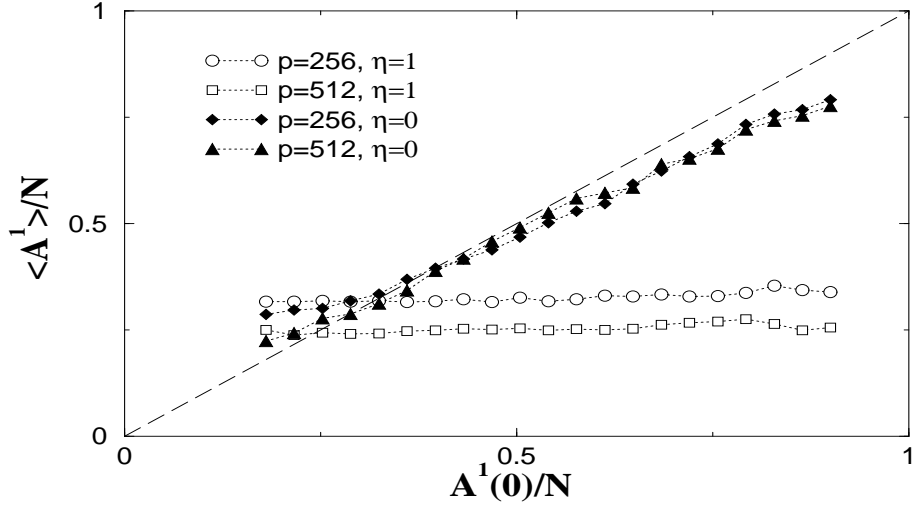}
\caption{Same as fig. \ref{corrup} but in the spin glass phase
($\alpha=1$ and $g=0.1$).}\label{corrupa1}
\end{figure}

Fig. \ref{corrupa1} shows that a very different scenario takes place
in the spin glass phase. While for $\eta=1$ even starting from maximal
overlap the dynamics converges to a spin glass state ($\langle
A^1\rangle /N\to 0$ as $1/\sqrt{N}$ when $N\to\infty$), for $\eta=0$
the stationary state preserves the initial overlap. In order to
understand this behaviour it is useful to compute the number of
stationary states of the majority game as a function of the parameters
$\alpha,~g$ and $\eta$.

\section{Number of stationary states}

In order to learn more about the phase space of the majority game we
calculate the number of stationary states as a function of various
parameters.  The stability relation (\ref{stab}) can be written in the
form:

\be
m_i\overline{\xi_iA}-\eta\overline{\xi_i^2}>0,~~~~m_i=\pm 1\label{stab2}
\ee

\noindent The average number of stationary states can be defined as:

\be
\Phi=\lda\sum_{\{m_i=\pm 1\}}\prod_{i=1}^N\Theta\lb(m_i\overline{\xi_iA}
-\eta\overline{\xi_i^2}\rb)\rda~,
\ee

\noindent where $\lda ...\rda$ stands for the average over random patterns 
$\xi_i^\mu$. This quantity is not self averaging and diverges exponentially 
with $N$. Therefore we will calculate the annealed entropy:

\be
s_a=\frac{1}{N}S_a=\frac{1}{N}\ln\Phi
\ee

Since the average over the disorder is inside the logarithm (annealed 
approximation) the calculations are straightforward (see 
\cite{matteo_ns,gardner}) and lead to the following result:

\be
s_a=
\max_{c,\hat{c},\Gamma,\hat{\Gamma},\gamma,\hat{\gamma}}
\lb\{ s_a(c,\hat{c},\Gamma,\hat{\Gamma},\gamma,\hat{\gamma})
\rb\}
\ee

\bea \fl s_a(c,\hat{c},\Gamma,\hat{\Gamma},\gamma,\hat{\gamma})&=&
c\hat{c}-\alpha\gamma\hat{\gamma}
+\alpha^2\Gamma\hat{\Gamma}-\fr{\alpha}{2}\ln\lb[2\Gamma+(\gamma-1)^2\rb]\nn
\\
\fl &+&\ln\lb[\cosh(\hat{c})-\fr{1}{2}(1-g)\lb
[\e^{-\hat{c}}\erf\lb(\fr{2(1-g)(\eta-\hat{\gamma})
-\fr{2}{\alpha}c}{2\sqrt{2(1-g)\hat{\Gamma}}}\rb)\rb.\rb.\nn
\\
\fl &+&\lb.\lb.\e^{\hat{c}}\erf\lb(\fr{2(1-g)(\eta-\hat{\gamma})
+\fr{2}{\alpha}c}{2\sqrt{2(1-g)\hat{\Gamma}}}\rb)
\rb]
-g\cosh(\hat{c})\erf\lb(\fr{2(1-g)(\eta-\hat{\gamma})}
{2\sqrt{2(1-g)\hat{\Gamma}}}\rb)\rb]\nn\\
\label{entropy}
\eea

\begin{figure}[h]
\vspace{1cm}
\includegraphics[width=13cm,height=7cm,angle=0]{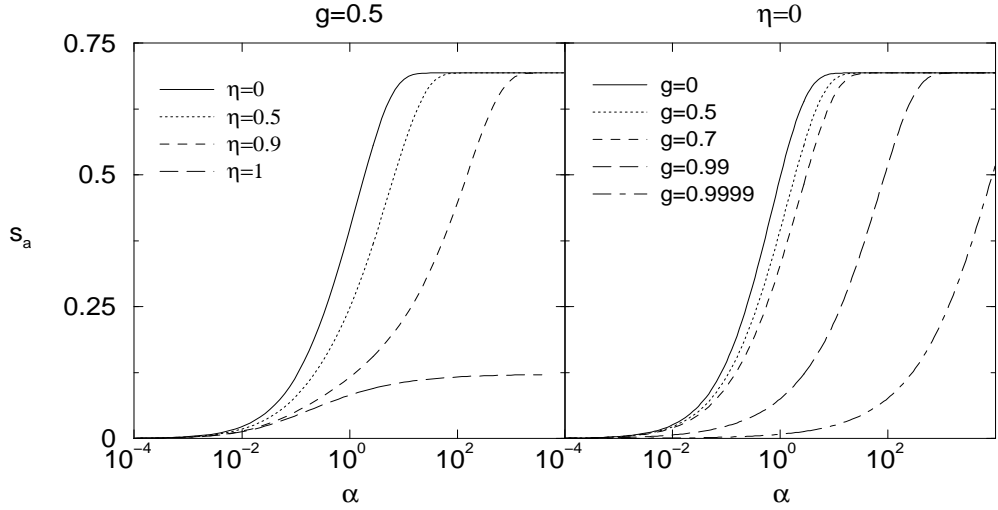}
\caption{$\alpha$ dependence of entropy $s_a$ for various values of the
parameters $g$ and $\eta$}\label{entplot}
\end{figure}

In fig. \ref{entplot} the annealed entropy is plotted as a function 
of $g$, $\eta$ and $\alpha$.

\begin{figure}[h]
\includegraphics[width=10cm,height=7cm,angle=0]{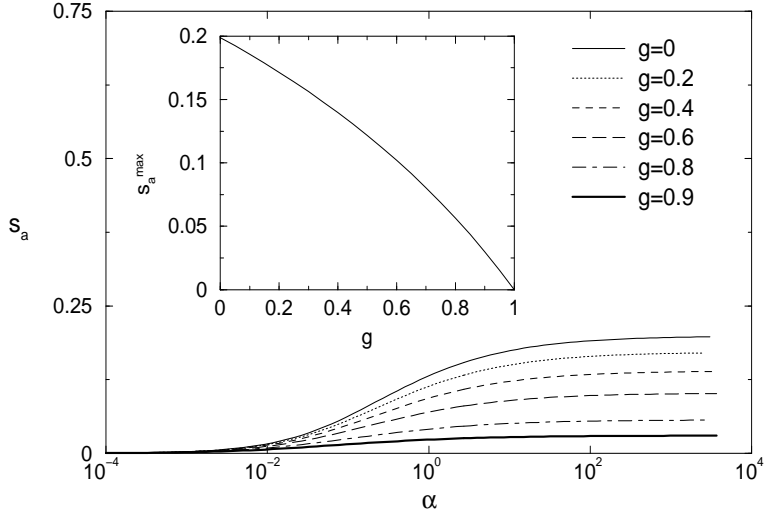}
\caption{$\alpha$ dependence of entropy $s_a$ for $\eta=1$ and various 
values of $g$. Inset: Maximal entropy $s_a^{max}$ as a function of $g$ 
for $\eta=1$ and $\alpha\to\infty$.}\label{sasat}
\end{figure}

For large values of $\alpha$ the number of stationary states increases 
dramatically when $\eta$ decreases from $1$ to $0$ . 
As $\alpha$ grows $s_a$ saturates at $\ln (2)$ for $\eta\neq 1$. 
This can be understood observing that when
\[
\left|\overline{\xi_i A_{-i}}\right|\equiv\left|\overline{\Omega\xi_i}
+\sum_{j\neq i}\overline{\xi_i\xi_j}m_j\right|<(1-\eta)\overline{\xi_i^2}
\simeq (1-\eta)(1-g)
\]
eq. (\ref{stab}) is satisfied for both $m_i=\pm 1$.  In words, when
the effective field $A_{-i}$ on spin $i$ due to the other spins $m_j$
is weak enough (i.e. its absolute value is smaller than
$(1-g)(1-\eta)$), $m_i$ can take both values. If it happens for every
$i$ all the states of the system (expressed in terms of configuration
$\{m_i\}$) are stationary and thus $s_a=\ln(2)$.  For large $\alpha$
the effective fields become small in absolute value. Indeed
$\overline{\xi_i A_{-i}}$ is well approximated by a Gaussian variable
with zero mean and variance $2|{\cal H}_0|/p\sim 1/\alpha$. As a
result $s_a\to\ln (2)$ as $\alpha\to \infty$. Only when $\eta=1$ or $g=1$
$s_a$ saturates to values smaller than $\ln(2)$ (see
fig. \ref{sasat}).

The strong $\eta$ dependence of $s_a$ displayed in the first plot in
fig.  \ref{entplot} explains the difference in the dynamical
behaviours of the model in the spin glass phase with $\eta=0$ and
$\eta=1$ (fig. \ref{corrupa1}).  Since for $\eta=0$ the stationary
states are very dense in the phase space, the system does not move far
from the initial state before it is trapped in one of the fixed points
of the dynamics. Thus the initial overlap changes very little. In the
case $\eta=1$ the number of stationary states is much smaller and the
system goes far away from the initial state (the initial non-zero
overlap vanishes for $N\to\infty$). It is important to remark that
the states with a non-zero overlap which are stable for $\eta=0$ in the
spin glass phase are not attractors in the usual sense because their
basin of attraction vanishes in the thermodynamic limit.

\begin{figure}[h]
\begin{center}
\includegraphics[width=12cm,height=6cm,angle=0]{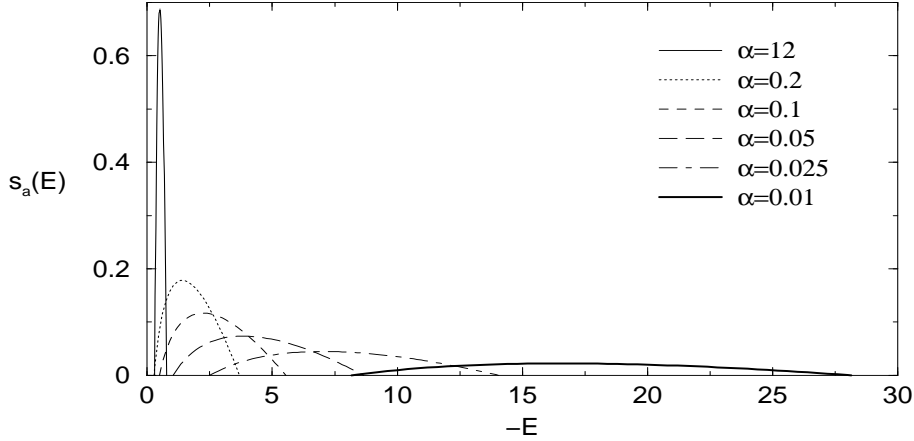}
\end{center}
\caption{Entropy $s_a(E)$ as a function of energy for $g=0.5$, $\eta=0$ and 
various values of $\alpha$}\label{entren}
\end{figure}

\begin{figure}[h]
\includegraphics[width=13cm,height=6cm,angle=0]{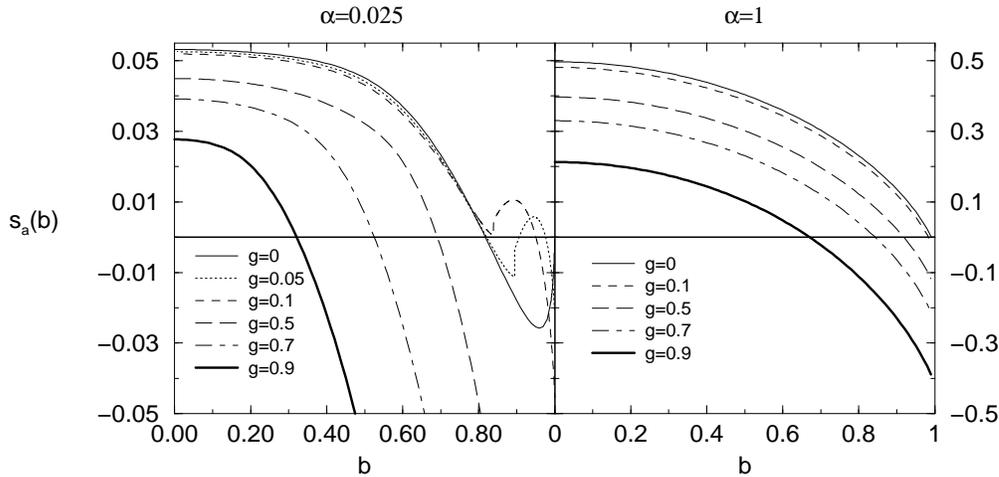}
\caption{Entropy $s_a(b)$ as a function of the overlap for $\eta=0$ and 
various values of $g$}\label{entrover1}
\end{figure}

It is easy to find $s_a$ as a function of Energy $E$:
\be
s_a(E)=
\max_{c,\hat{c},\Gamma,\hat{\Gamma},\gamma,\hat{\gamma},u}
\lb\{ s_a(c,\hat{c},\Gamma,\hat{\Gamma},\gamma,\hat{\gamma})+
u E +\frac{1}{2}\frac{u}{\alpha}c^2
\rb\}
\ee
or overlap $b$.
\be
s_a(b)=
\max_{c,\hat{c},\Gamma,\hat{\Gamma},\gamma,\hat{\gamma},u}
\lb\{ s_a(c,\hat{c},\Gamma,\hat{\Gamma},\gamma,\hat{\gamma})+
u b-u c
\rb\}
\ee
The results are presented in figs. \ref{entren} and \ref{entrover1}.

The energy dependence of $s_a$ enables us to determine the average energy of the
infinite system (see the curve on fig. \ref{encom}) and explains the size of the
error bars of the simulation results. The larger error bars are due to the wider
distribution of the energy (smaller $\alpha$).

The $b$ dependence of $s_a$ deep in the retrieval phase has a
different character than in the spin glass phase (see
fig. \ref{entrover1}).
The gap (region where $s_a(b)<0$) in the distribution in fig.
\ref{entrover1} disappears close to the phase boundary. 
We suppose that in reality the gap vanishes precisely at this boundary 
(compare \cite{gardner}).
Unfortunately due to the inadequacy of the annealed approximation we 
are not in a position to draw more quantitative conclusions.

This inadequacy of the annealed approximation used
to calculate the entropy $s_a$ is much more evident than
in the case of the minority game with $\eta=1$ \cite{matteo_ns}.
One can see it by comparing the maximal allowed value of the
overlap $b_{max}=1-g$ with the maximal value of $b$ for which
$s_a(b)>0$ (fig. \ref{entrover1}). For all $g>0$ the function $s_a(b)$
suggests the existence of stationary states with $b>b_{max}$.  Also
the discrepancy between the energy profiles (fig. \ref{encom}) should
be attributed to the use of the annealed approximation, which indeed
overestimates the real quantities.

We tried to calculate the quenched entropy using replica method (see
\cite{braymoore}), but up to now we were not able to solve the arising
numerical problems.

\section{Conclusions and discussion}

We have shown that the stationary states of the majority game
correspond to the local minima of a Hopfield type hamiltonian and are
attained when all agents ``freeze'', i.e. use always the same
strategy. Stationary states are not necessarily Nash equilibria except
when agents correctly account for their impact on the aggregate
($\eta=1$). Depending on the parameters, the system can be in one of
two phases: a retrieval phase characterised by attractors with a
macroscopic overlap $A^1\sim O(N)$ and a spin glass phase with no
retrieval. A macroscopic overlap can also be sustained, in the spin
glass phase, for $\eta$ small. We attribute this phenomenon to the
self-reinforcing term $(1-\eta)\overline{\xi_i^2}s_i$ in the dynamics
which causes a dramatic increase in the number of stationary states as
$\eta$ decreases. These results extend to the on-line version of the
game. Indeed the equations for the stationary states are the same and,
since agents freeze in the long run, fluctuations play no role
(contrary to the case of the minority game~\cite{matteo_con}).

These results allow us to draw some suggestions on the behaviour of
systems of interacting agents driven by conformity or by increasing
returns. The occurrence of a macroscopic overlap $A^1\sim O(N)$ may
correspond to crowd effects such as fashions and trends, when a large
fraction of agents behave similarly in some respect, or to economic
concentration, when, for example, one particular place is arbitrarily
selected for large scale investments. The development of these crowd
effects requires: {\em i)} that the number of agents is large compared
to the number of resources ($\alpha$ small), {\em ii)} a sufficient
differentiation between strategies of agents ($g<2/3$) and {\em iii)}
a large enough initial bias (i.e. an initial macroscopic overlap)
towards a particular resource, fashion or place.  Finally crowd
effects can be sustained under more general conditions (i.e. in the
spin glass phase) if agents do not behave strategically, i.e. if they
neglect their impact on the aggregate ($\eta$ small).

Besides the relevance of the model as a system of heterogeneous
interacting agents, it is also interesting as an example of
non-Glauber dynamics in the energy landscape of Hopfield type
hamiltonians. It is remarkable that, in spite of the fact that the
dynamics of $y_i$ does not satisfy detailed balance, the statistical
mechanics picture remains quite accurate.

\section*{Acknowledgements}

This work was done within the EU research and training network STIPCO
under the contract: STIPCO HPRN-CT-2002-00319.

\end{document}